\documentclass{aa}
\usepackage[varg]{txfonts}
\usepackage{natbib}
\usepackage{multirow}
\usepackage{graphicx}
\usepackage{times}
\usepackage{color}
\usepackage{siunitx}
\usepackage{float}
\usepackage{tikz}
\bibpunct{(}{)}{;}{a}{}{,}

\newcommand{\hd}{HD~135160 }
\newcommand{\hk}{HD~135160}

\newcommand{\respefoe}{{\tt reSPEFO}}

\newcommand{\fotel}{{\tt FOTEL} }
\newcommand{\fotele}{{\tt FOTEL}}
\newcommand{\korel}{{\tt KOREL} }
\newcommand{\korele}{{\tt KOREL}}

\newcommand{\arcm}{$^\prime$}
\newcommand{\arcs}{$^{\prime\prime}$}
\newcommand{\m}{$^{\rm m}\!\!.$}

\newcommand{\ks}{km~s$^{-1}$}

\newcommand{\tef}{$T_{\rm eff}$ }

\newcommand{\cd}{c$\,$d$^{-1}$}

\newcommand{\ha}{H$\alpha$ }

\newcommand{\hb}{H$\beta$ }

\newcommand{\hae}{H$\alpha$}

\begin{document}

\makeatletter
\let\switchlinenumbers\relax
\let\runningpagewiselinenumbers\relax
\let\pagewiselinenumbers\relax
\let\linenumbers\relax
\let\nolinenumbers\relax
\makeatother

  \title{Quadruple system HD~135160 in a unique 2+2 configuration}
  \author{M.~Zummer\inst{1}\and
          P.~Harmanec\inst{1}\and
          B.~Barlow\inst{2,3}\and
          M.~Blackford\inst{4}\and
      J.~Švrčková\inst{1,5}
}
   \offprints{M. Zummer\,\\
   \email \ michalmiskozummer@gmail.com}

  \institute{
   Astronomical Institute of Charles University,
   Faculty of Mathematics and Physics,\hfill\break
   V~Hole\v{s}ovi\v{c}k\'ach~2, CZ-180 00 Praha~8 - Troja, Czech Republic
 \and
Department of Physics and Astronomy, University of North Carolina at Chapel Hill,
Chapel Hill, NC, 27599, USA
 \and
Department of Physics and Astronomy,
High Point University, High Point, NC, 27268, USA
\and
   Variable Stars South, Congarinni Observatory, Congarinni, NSW, Australia~2447
\and
European Organisation for Astronomical Research in the Southern Hemisphere (ESO), Casilla 19001, Santiago 19, Chile
}
\date{Received October 10, 2025, accepted November 08, 2025}

\abstract{ 
Analysing a large body of observational data, we found that HD~135160 is a~quadruple 2+2 system, composed of a massive ellipsoidal binary ('heartbeat' star) with components Aa and Ab in an eccentric 8.234~d orbit and an~eclipsing binary (with components Ba and Bb), with a~5.853~d period and partial eclipses that have already been reported from the space photometry secured by the Transiting Exoplanet Survey Satellite (TESS). Our systematic echelle spectroscopy, secured since September 2021, led to the discovery that the optical spectra are dominated by spectral lines of three early-type stars, two moving around each other on a 8.234~d orbit of a high eccentricity, which causes periodic brightenings near the periastron passage, and the third one (component Ba) being the brighter component of the 5.853~d binary. Both pairs are physically bounded and revolve around each other with a~period somewhere between 1600 and 2200 days (4.4 -- 6 years). The object exhibits small cyclic light variations of a~variable amplitude and characteristic time scale of 0\fd071 (14.14~\cd), seen throughout the whole orbit. The nature of these tiny changes deserves further investigation. It also seems that the earlier classification of the object as a Be star is unfounded.
}


\authorrunning{M.~Zummer et al.}
\titlerunning{Quadruple system HD~135160}
\maketitle

\section {Introduction}

Due to tidal distortion of the components, eccentric binaries exhibit periodic brightenings near the periastron passage. As a matter of curiosity, we note that probably the first case, when mild light brightening was observed near the periastron passage in a binary with an eccentric orbit, is the study of 96~Her (HD~164852) based on radial velocities (RVs) from photographic spectra and ground-based $UBV$ photometry
from Hvar \citep{koubsky85}. An attempt was made to model the effect by \citet{hadrava86}, and in relation to binary pulsars by \citet{Kumar1995ApJ}. This class of binaries was recognised only much later, using the Kepler space telescope data by \citet{Welsh_2011} and dubbed 'heartbeat stars' by \citet{Thompson_2012}. Initially, mostly low-mass stars ($M\leq2  \mathcal{M}_\odot$) were found \citep{Shporer_2016}, while later a sample of massive heartbeat stars from Transiting Exoplanet Survey Satellite (TESS) was found and analysed by \citet{KolaczekSzymanski2021}.

\citet{Pejcha2013} predicted that quadruple systems in a 2+2 configuration may be very common among heartbeat stars and that their formation is more prevalent than that of triples or binaries. Although some massive heartbeat multiple systems have been observed, we are unaware of any study of a quadruple system with the 2+2 configuration with a heartbeat binary plus an eclipsing binary, as is the case in this study.

Only a handful of studies are devoted to massive heartbeat stars, for example \citet{Pablo2017} studied O9 III + B1 III/IV binary $\iota$ Ori and found multiple tidally excited oscillations (TEOs) in an O type star. \citet{MacLeod2023} studied the blue supergiant MACHO 80.7443.1718 with the most extreme heartbeat;  \citet{SzymanskiMACHO_2022, SzymanskiMACHO_2024} noted that its light variation is unlikely to be caused by an extreme tidal distortion and that the ellipsoidal distortion may be of secondary importance in shaping the light curve.

\hd is a bright ($V=5$\m8) B0.5V-B1V star (also known as HR~5661,
CPD$-60^\circ$5698, HIP~74750, TIC~455463415 or MWC 234;
$\alpha_{2000.0}$ = 15$^{\rm h}$16$^{\rm m}$36.$\!\!^{\rm s}$69,
$\delta_{2000.0}$ = $-60^\circ$54\arcm14\farcs4) and is the brightest member
of a visual binary system WDS~J15166-6054A, with 8\m55 component B
at a separation of 1\farcs2, and 11\m62 component C at a slowly varying
separation of 20\arcs\ to 27\arcs\ over more than a century. It has also been
reported to be a Be star by \citet{payne27}. \citet{thack69} published two
radial velocities (RVs) and argued that \hd shares a common space motion with
$\delta$~Cir. In their survey of the multiplicity of high-mass stars, 
\citet{chini2012} classified \hd as a~spectroscopic binary with two
visible spectra, based on the two spectrograms they had.

Our interest in this star started with an email that one of us, Mark Blackford, sent to Petr Harmanec on 15 August 2021.
He was looking for a suitable comparison star for his planned photometry of $\delta$~Cir and looked at the TESS photometry
of HD~135160. He found that \hd is an eclipsing binary with a period of 5\fd853 and gravitational distortion of one star, and noted the
presence of humps that did not correlate with the eclipse period and that small oscillations are also present throughout the orbit.

The eclipsing binary was officially reported in the paper by \citet{ij2021}, which was devoted to an overview of the OBA-type binaries observed by TESS. These authors identified \hd as an~eclipsing binary with a~period 5\fd84360, having a~superior conjunction at BJD~2458633.01530, with the depths of the primary
and secondary minima of 0\m0217 and 0\m0062, respectively.
\citet{prsa2022} published a catalogue of 4584 eclipsing binaries discovered in TESS data from sectors 1 to 26. For \hd they quote a somewhat different
value of the period, 5\fd85205(62), and the time of the first primary eclipse
of BJD~2458627.15570(62). They estimated the depths of the primary and secondary 
minima as 0\m022 and 0\m004, respectively.

In the hope of obtaining an orbital solution for the eclipsing binary and determining the system properties, we started systematic spectroscopic observations of the object
with the \mbox{CHIRON} echelle spectrograph at the Cerro Tololo Inter-American Observatory (CTIO). 
In this paper, we report the surprising results of our effort.

\section {Observations and data reductions}
Throughout this paper, we give all times of observations using
the reduced heliocentric Julian dates defined by
    $\mathrm{RJD}=\mathrm{HJD} - 2400000.0\,.$
We denote $T_\textrm{0, perpass}$ for the time of periastron passage and $T_\textrm{0, supconj}$ for the time of superior conjunction.

\subsection{Spectroscopy}
We obtained 70 high-resolution echelle spectra of \hd from the CHIRON spectrograph at the CTIO over a~time interval RJD~59472.4654 -- 60916.4865, with a spectral resolution of $R\sim25000$ and exposure times of 30~s, 100~s, 300~s, and 500~s in the spectral range 450 - 890 nm. Their overview is given in Table~\ref{jouspe}. Although a small number of echelle spectra are available in online databases, they could not be easily combined with our data, and we have decided not to use them. All spectra were converted to 1D images and wavelength-calibrated with the pipeline of the CTIO Observatory. All consecutive reductions, i.e. rectification, removal of cosmic-ray spikes and occasional flaws, and RV measurements based on the tracking paper method, were carried out with the latest version 2 of the program \texttt{reSPEFO}\footnote{\url{https://astro.troja.mff.cuni.cz/projects/respefo/}}. 

During the initial inspection of the spectra, up to three different components were found in the hydrogen and helium lines. In the two strongest \ion{He}{ii} lines at 4686 and 5411~\AA, a spectral line of only one component is visible, while two or even three lines were seen in the \ion {He}{i} lines at 4923, 5016, and 6678~\AA, \hae, and \hb. For all these lines, RVs were independently measured by Michal Zummer and Petr Harmanec, and mean values were adopted with errors being the standard deviation. They are tabulated in detail in Table~2, only available at the Strasbourg Astronomical Data Centre (CDS).

\begin{table}[h]
\begin{center}
\caption[]{Journal of CHIRON spectra secured during multiple intervals.}\label{jouspe}
\begin{tabular}{crcrl}
\hline\hline\noalign{\smallskip}
Time interval&No.    &Wavelength &Spectral  \\
             &of     & range     &res.\\
    (HJD-2400000)&spectra& (\AA)   \\
\noalign{\smallskip}\hline\noalign{\smallskip}
 59472.4654 -- 59497.4836& 10&\multirow{7}{*}{4500--8900}&\multirow{7}{*}{25000}\\
 59620.8824 -- 59625.8808& 23&\\
 59819.4608 -- 59820.4618& 6 &\\
 60124.6072 -- 60130.7476& 14&\\
 60727.8972 -- 60738.9061& 4&\\
 60783.7428 -- 60797.6659& 8&\\
 60905.4584 -- 60916.4865& 5&\\
 
\noalign{\smallskip}\hline\noalign{\smallskip}
\end{tabular}
\end{center}
\end{table}

\subsection{\textup{TESS} photometry}
Although primarily designed for the monitoring of transiting exoplanets, TESS has been invaluable in obtaining a continuous series of very accurate observations over about a month for a given area in the sky. This provided densely covered and precise light curves of eclipsing and ellipsoidal binaries, including the discoveries of new ones. We have extracted data using the \texttt{lightcurve} \texttt{python} library\footnote{\url{https://lightkurve.github.io/lightkurve/}} \citep{Lightkurve2018} from TESS sectors 12 and 65, using the Simple Aperture Photometry flux (SAP) to avoid possible removal of real astrophysical variability by detrending. To facilitate further analysis, we have normalised the light curves (LCs) of both available sectors separately to the median values.

\section{Towards orbital solutions}
\begin{figure}[h]
\resizebox{\hsize}{!}{\includegraphics[angle=0]{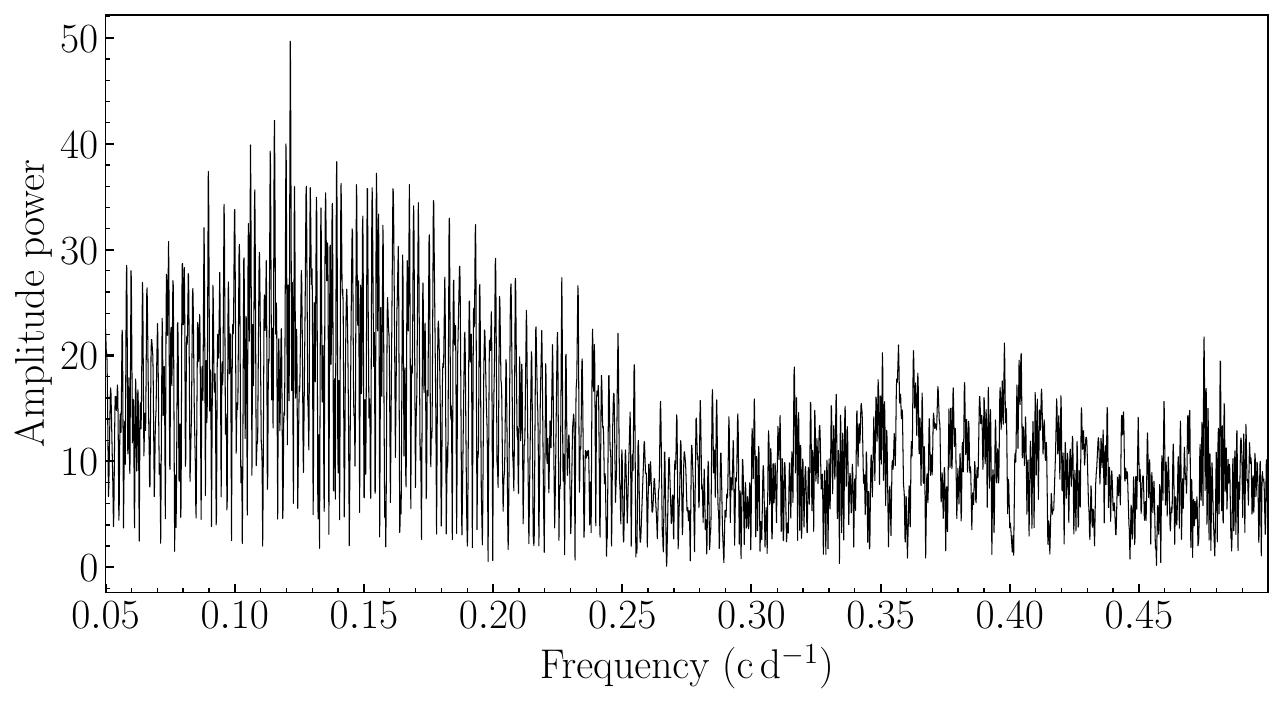}}
\caption{Amplitude periodogram of RVs measured on \ion{He}{ii}~4686~\AA\ line. The highest peak at frequency of 0.1214~\cd\ corresponds to period of 8\fd234.}
\label{periodogram}
\end{figure}

\begin{figure}[h]
\resizebox{\hsize}{!}{\includegraphics[angle=0]{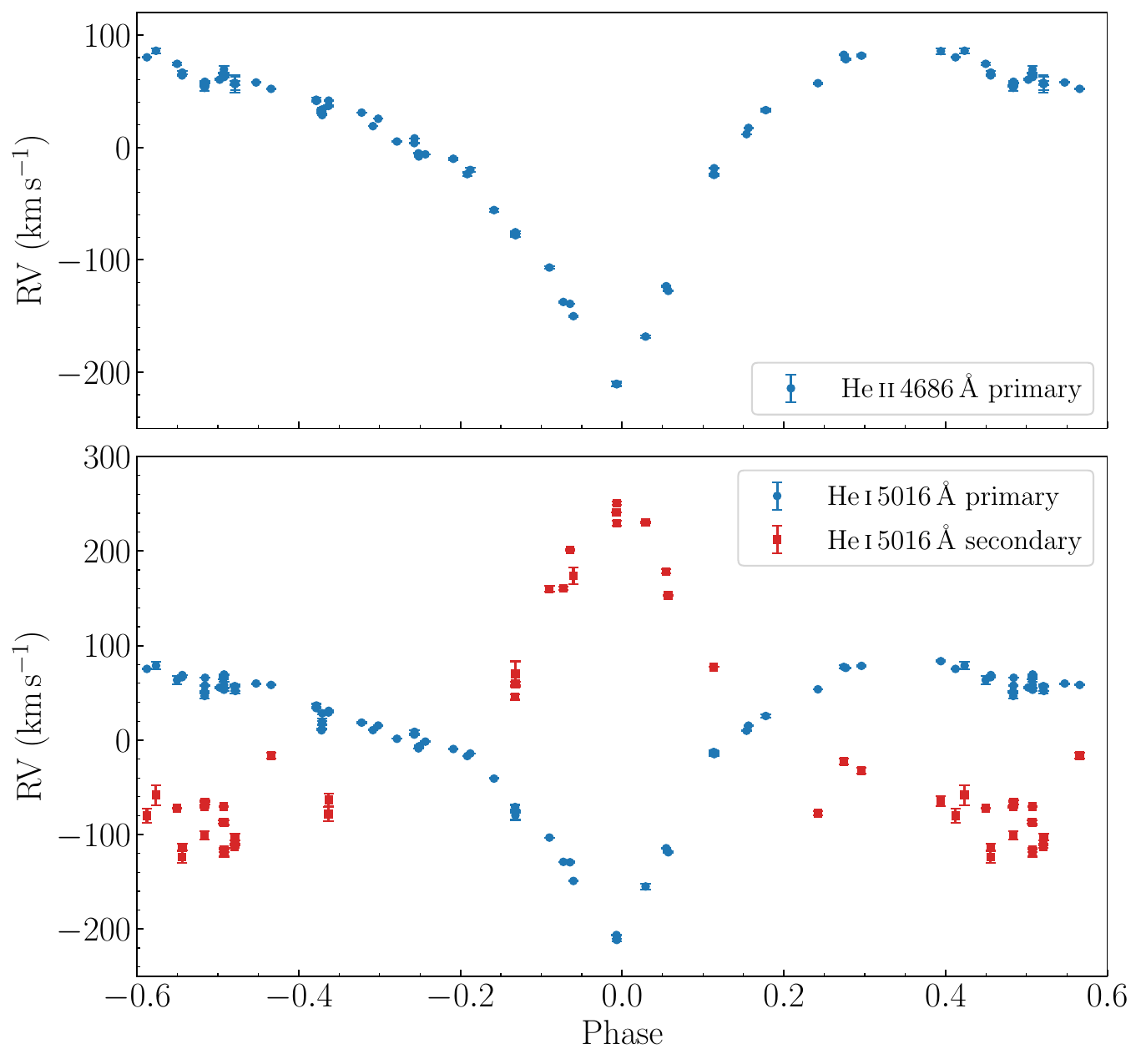}}
\caption{Top: RV curve of the 8\fd234 primary measured in {\tt reSPEFO} on \ion{He}{ii}~4686~\AA\ line.
Bottom: RV curves of the 8\fd234 primary (blue) and secondary (red) based on the {\tt reSPEFO} RVs of the clean \ion{He}{i}~5016~\AA\ 
line. A preliminary ephemeris $T_{\rm RVmin.}$ = RJD $60003.740 + 8\fd2347\times E$ was used in both plots. Errors are taken as a mean of two independent measurements and are too small to be seen.} 
\label{rvhe}
\end{figure}

\begin{figure}[h]
\resizebox{\hsize}{!}{\includegraphics[angle=0]{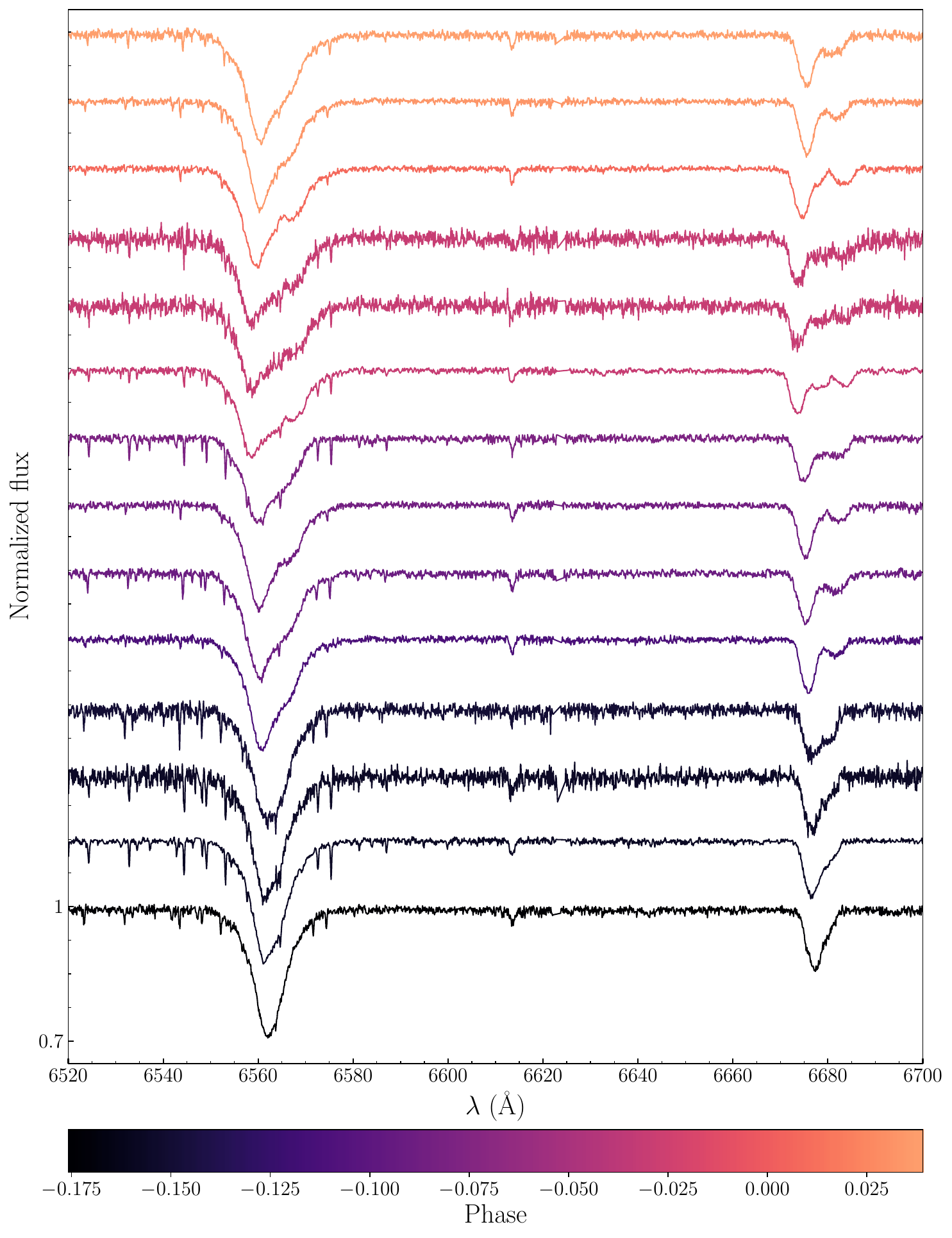}}
\caption{Parts of the CHIRON spectra 
with H$\alpha$ and \ion{He}{i}~6678~\AA\ ordered according to their phase for ephemeris $T_\textrm{0, perpass}=\textrm{RJD } 59995.6424 + 8.23459\times E $. Only some of the spectra are plotted, corresponding to around a third of the orbital period, so that change of the position of lines of 3 components (Aa, Ab, and Ba) is visible. Distances between spectra do not correspond to the distance in phase. Difference in $S/N$ can be seen between 30 s, 300 s, and 500 s exposures.}
\label{HalphaHe6678_phased}
\end{figure}

The plots of the RVs measured versus the phase of the 5\fd853 period indicate that they do not vary with that period.
This prompted us to carry out a~period search using a program based on the \citet{deeming75} method for amplitude periodograms, for the RVs of the \ion{He}{ii}~4686~\AA\ line, which appeared single in all our spectrograms.
The corresponding periodogram, shown in Fig. \ref{periodogram}, clearly indicates a~new period of 
8\fd234 (0.1214~\cd). We also plot the RV curves based on the measurements in \texttt{reSPEFO} in Fig.~\ref{rvhe}.
Segments of the spectra containing the H$\alpha$ and \ion{He}{i}~ 6678 \AA\ lines, ordered according to the phase of the 8\fd234 period, are shown in Fig. \ref{HalphaHe6678_phased} with up to 3 distinguishable stellar components. 

\subsection{Quadruple system}

\begin{figure}[h]
\centering
\begin{tikzpicture}
    \draw (0,0) circle (0.5);      
    \draw (3,0) circle (0.5);        
    \node at (0,-1) {Aa};
    \node at (3,-1) {Ab};

    \draw (5,0) circle (0.5);        
    \draw (8,0) circle (0.4);        
    \node at (5,-1) {Ba};
    \node at (8,-1) {Bb};

    \draw (0,0.5) -- (0,1) -- (3,1) -- (3,0.5);

    \draw (5,0.5) -- (5,1) -- (8,1) -- (8,0.4);

    \draw (1.5,1) -- (1.5,2) -- (6.5,2) -- (6.5,1);

    \node at (1.5,0.7) {$P_{\scriptscriptstyle\mathrm{A}}=8.234$ d};
    \node at (6.5,0.7) {$P_{\scriptscriptstyle\mathrm{B}}=5.853$ d};
    \node at (4,1.5) {$P_{\scriptscriptstyle\mathrm{AB}}\approx4.4 - 6$ years};

    \node at (1.5,-1.5) {heartbeat};
    \node at (6.5,-1.5) {eclipsing};
\end{tikzpicture}

\caption{Scheme of the HD~135160 system based on this study.}
\label{scheme}
\end{figure}
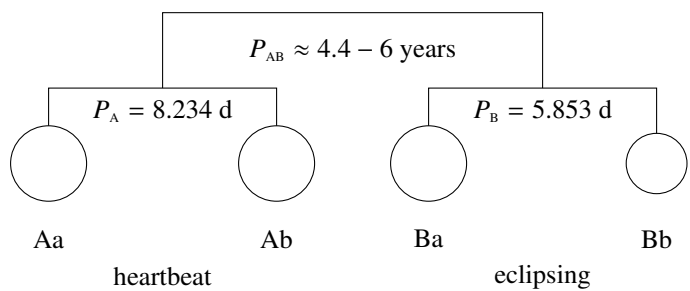

While a triple star system has only one stable configuration 2+1, quadruples are stable in 3+1 and 2+2. As two close periods of 5\fd853 and 8\fd234 are present, we have concluded from the argument of dynamical stability that \hd consists of 2 binaries. A schematic
illustration of the multiple system HD~135160, based on current knowledge, can be seen in Fig. \ref{scheme}. Only three of four stellar components are visible in the spectra, corresponding to early-type stars Aa, Ab, and Ba.

\subsection{Preliminary orbital elements based on RV measurements}
\begin{figure}[h]
\centering
\resizebox{\hsize}{!}{\includegraphics[angle=0]{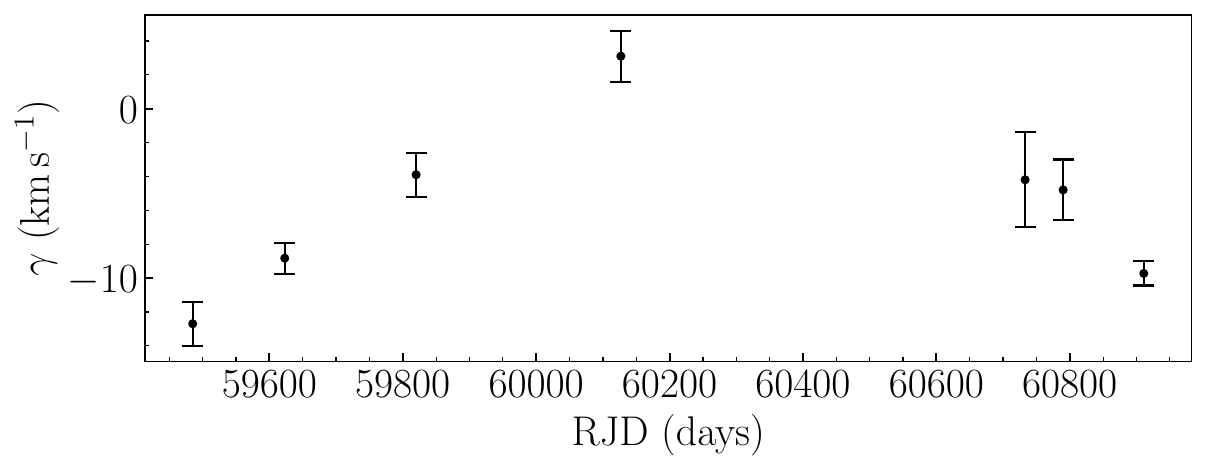}}
\caption{Locally derived systemic ($\gamma$) velocities of binary A, based on the orbital solution with \texttt{FOTEL} from RVs measured on \ion{He}{ii} 4686 \AA\ line.}
\label{gamas}
\end{figure}

To obtain the input values of the orbital elements of the 8\fd234 subsystem for a more sophisticated treatment, we derived trial orbital solutions using program \texttt{FOTEL} \citep{Hadrava1990,Hadrava2004_FOTEL} for the RVs, measured in \texttt{reSPEFO} on the \ion{He}{ii}~4686 line 
and the \ion{He}{i}~5016~\AA\ line, which is clean of neighbouring blends. These preliminary orbital elements are shown in
Table~\ref{rvrespefo}. For the \ion{He}{ii}~4686 line, we allowed calculation of individual systematic $\gamma$ velocities for each series of observation (see journal of spectra in Table \ref{jouspe}), which resulted in a significant decrease of the rms errors. This solution is in the last column of Table~\ref{rvrespefo}. A plot of the local $\gamma$'s from the observing intervals versus time is shown in Fig.~\ref{gamas} and clearly documents the orbital motion of the Aa-Ab binary in a long orbit. Our data still do not cover one whole mutual orbit of subsystems A and B, but based on the period analysis and subsequent modelling of the long orbit, it seems probable that the value of this long period is between 1600 and 2200 days.

\addtocounter{table}{1}
\begin{table*}[h]
\begin{center}
\caption[]{Preliminary orbital elements calculated using \texttt{FOTEL}.}
\label{rvrespefo}
\begin{tabular}{cccccl}
\hline\hline\noalign{\smallskip}
   Element&\ion{He}{ii} 4686 \AA\ RVs&\ion{He}{i}~5016~\AA\ RVs  & \ion{He}{ii} 4686 \AA\ RVs, local gammas       \\
\noalign{\smallskip}\hline\noalign{\smallskip}
$P$ (d)     &8\fd23488(0.00018)&8\fd23452(0.00038)&8\fd23496(0.00013)    \\
$T_\textrm{0, perpass}$ (RJD) &59995.675(0.024)&59995.651(0.047)&59995.669(0.020)\\
$T_\textrm{RV max.}$ (RJD)    &59998.446 & 59998.587&59998.487\\ 
$e$       &0.4759(0.0084)&0.491(0.017)&0.4668(0.0061)\\
$\omega$ ($^\circ$) & 205.1(1.3)&201.1(2.9)&204.6(1.1)\\
$K_1$ (\ks) &138.8(1.8)&132.2(5.1)&139.6(1.3)\\
$K_1/K_2$   & --   &0.769(0.032)& -- \\
$K_2$ (\ks) & --   &171.8& -- \\
$\gamma$ (\ks) &$-5.60$(0.84)&$-3.2$(1.9)&--\\
$\gamma_1$ (\ks)& --&--&$-12.7$(1.3)\\
$\gamma_2$ (\ks)& --&--&$-8.83$(0.95)\\
$\gamma_3$ (\ks)& --&--&$-3.9$(1.3)\\
$\gamma_4$ (\ks)& --&--&$+3.1$(1.5)\\
$\gamma_5$ (\ks)& --&--&$-4.2$(2.8)\\
$\gamma_6$ (\ks)& --&--&$-4.8$(1.8)\\
$\gamma_7$ (\ks)& --&--&$-9.73$(0.71)\\
rms (\ks)   &6.66&18.9&4.37\\
\noalign{\smallskip}\hline\noalign{\smallskip}
\end{tabular}
\tablefoot{Solution for the 8\fd234 Aa+Ab orbit based on \mbox{\texttt{reSPEFO}} RVs. For \ion{He}{ii} 4686 \AA\, only RVs of the primary Aa are available, while for \ion{He}{i}~5016~\AA\ RVs of the primary Aa and secondary Ab are available.}
\end{center}
\end{table*}

In several spectra, we also observed faint lines that belong to the primary Ba of the eclipsing 5\fd853 subsystem, but
reliable RV measurements could not be obtained in \texttt{reSPEFO}.

\begin{figure*}[t]
\centering
    \centering
    \includegraphics[width=0.49\linewidth]{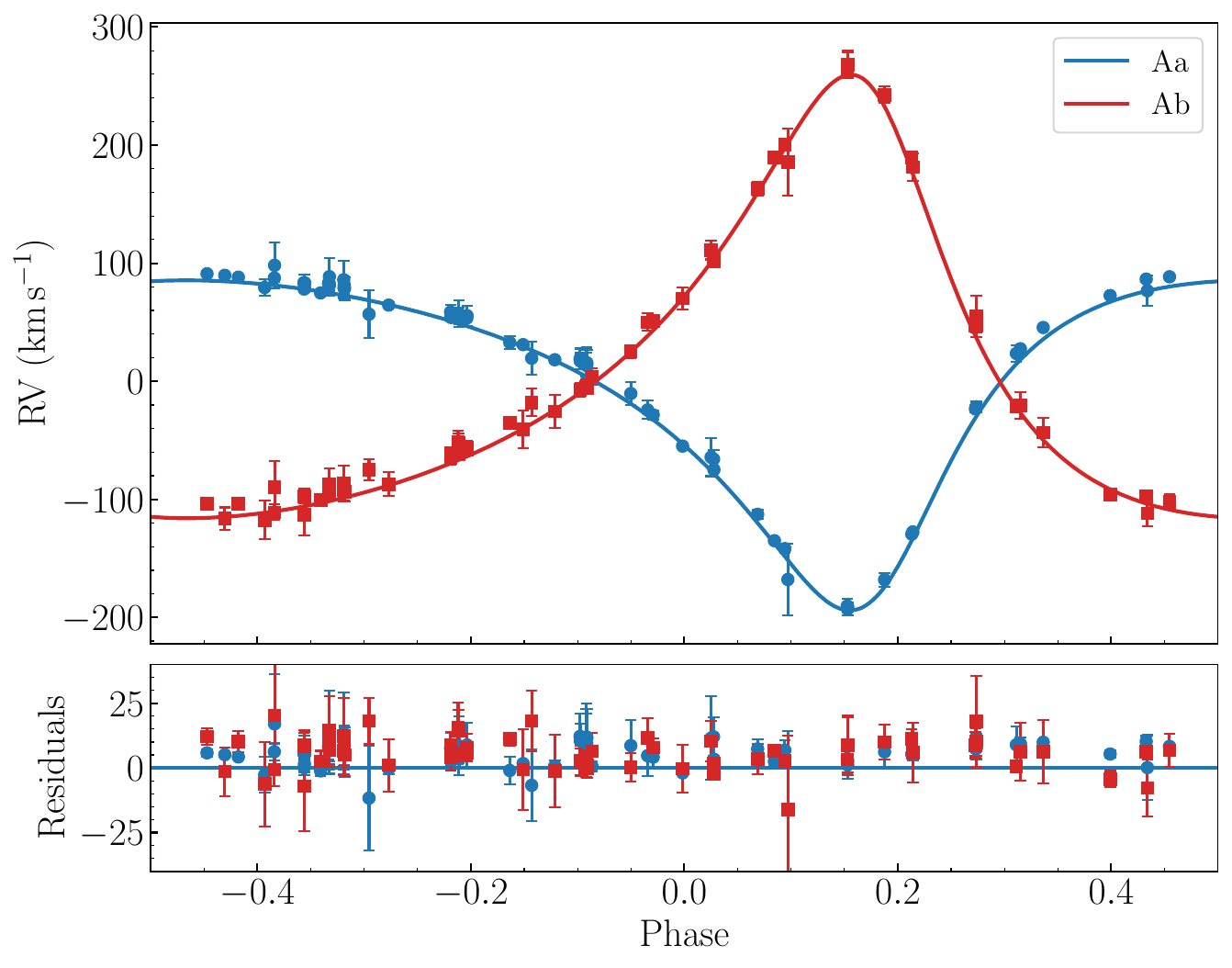}
    \includegraphics[width=0.49\linewidth]{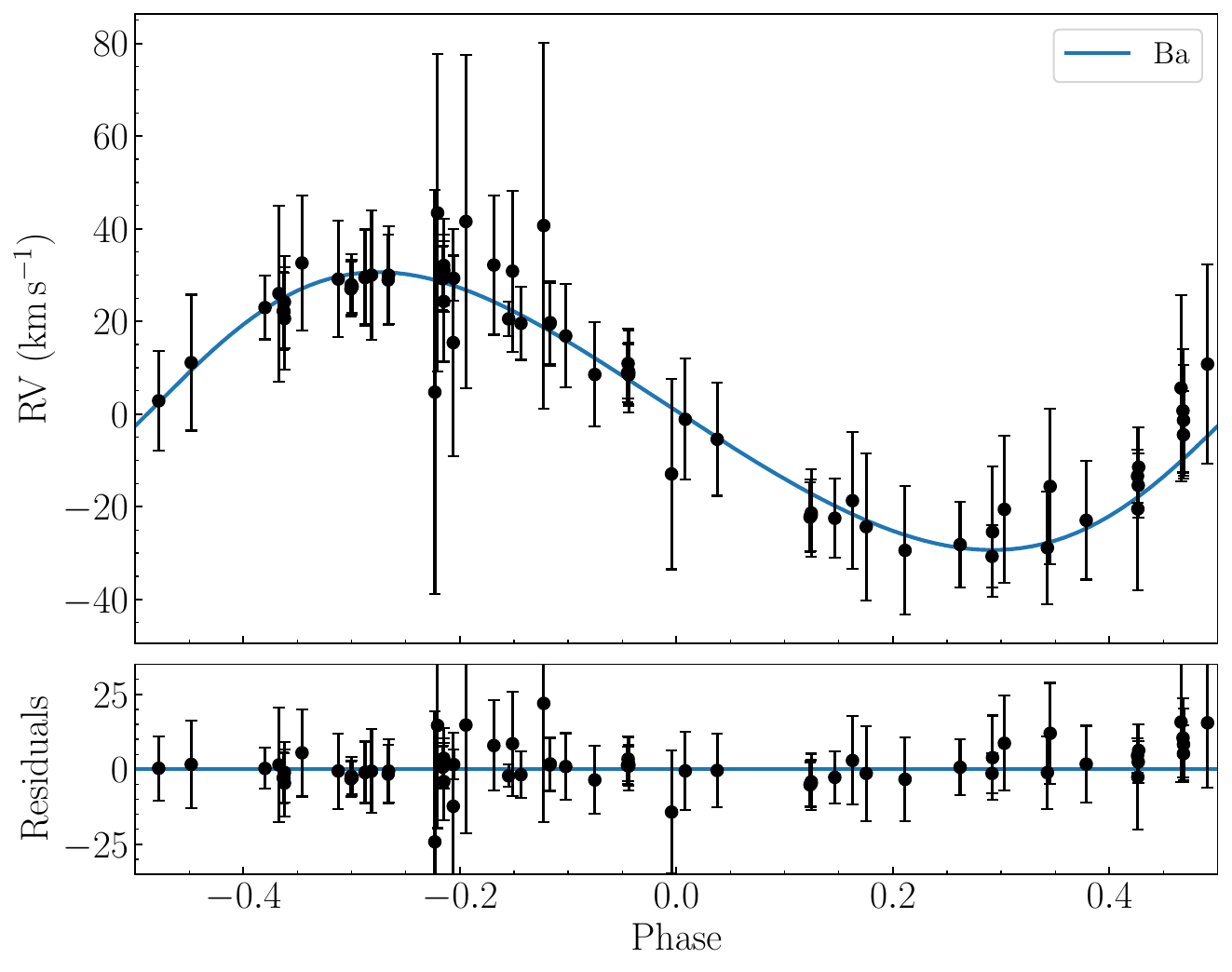}
    \caption{Left: \texttt{PHOEBE\,2} model of primary component Aa (blue) and secondary component Ab (red) of binary A, using RVs calculated from \texttt{KOREL} as a mean of \ion{He}{i} 4923, 5016, 6678 \AA\,, and H$\beta$. Ephemeris of $T_\textrm{0, perpass}=\textrm{RJD } 59995.64 + 8.23447\times E $ is used. Right: \texttt{PHOEBE\,2} model of primary component Ba of an  eclipsing binary B, using radial velocities calculated from \texttt{KOREL} as mean of \ion{He}{i} lines, H$\alpha$, and H$\beta$. Ephemeris of $T_\textrm{0, supconj}=\textrm{RJD } 58627.15 + 5.852864\times E$ is used. Error bars are shown as a standard deviation of RVs computed from selected lines. RVs are corrected for systemic gamma velocities calculated for each segment of observations.}
    \label{RV_3components}
\end{figure*}

\subsection{Spectra disentangling}
\begin{figure}[ht]
    \centering
    \includegraphics[width=\linewidth]{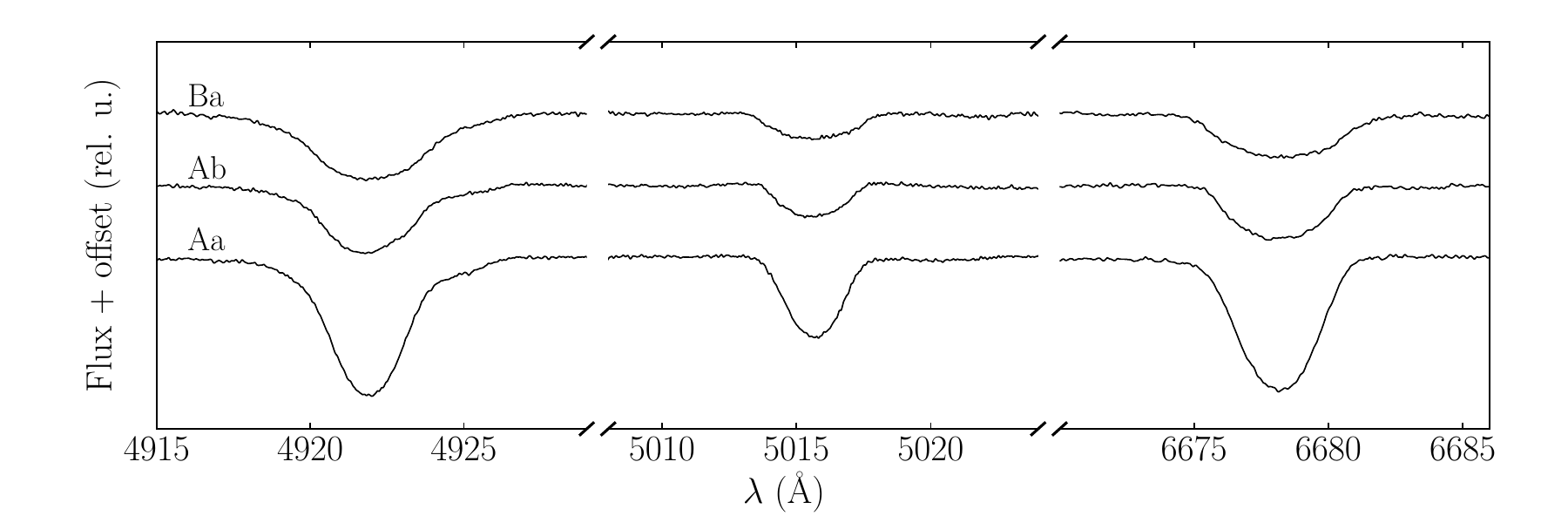}
    \caption{Examples of normalised disentangled components Aa, Ab, and Ba on lines \ion{He}{i} 4923, 5016, 6678~\AA.}
    \label{korel_spe}
\end{figure}

To obtain more reliable RVs for all three spectral components, we used the program \texttt{KOREL} for spectra disentangling \citep{Hadrava1995,Hadrava1997,Hadrava2004},
and derived solutions for both subsystems A and B on a long orbit in the neighbourhood of \ion{He}{i} 4923, 5016, 6678~\AA, H$\alpha$, and  H$\beta$ lines. We then calculated the mean RV,  the error being the standard deviation for components Aa and Ab from the \ion{He}{i} lines and H$\beta$, and with all of the above for component Ba. Different spectral lines were chosen for components Aa+Ab and Ba after multiple different trials, and the best suitable combination was chosen. The disentangled H$\alpha$ line caused greater errors in the RVs of components Aa and Ab, due to its large width, and was therefore omitted. The results for RVs can be seen in Fig.~\ref{RV_3components}. We re-ran calculations multiple times, allowing the program to fit the long orbit, with results that suggest a period between 1600 and 2200 days, as predicted from the period search of $\gamma$ velocities. Examples of disentangled components Aa, Ab, and Ba of the \ion{He}{i} lines are visible in Fig.~\ref{korel_spe}.

\subsection{Estimates of the radiative properties of system components}
\begin{figure}[h]
\resizebox{\hsize}{!}{\includegraphics[angle=0]{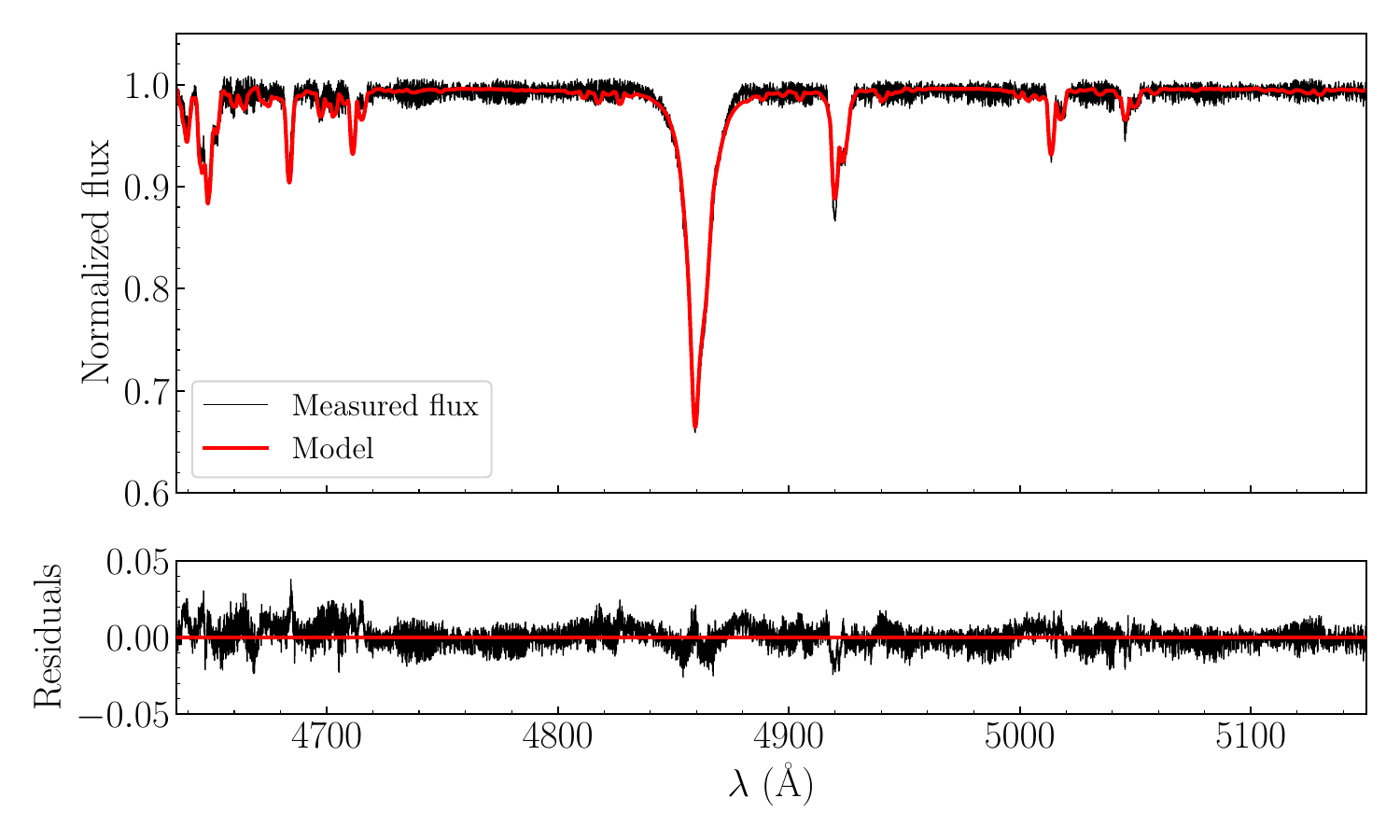}}
\caption{An example of  \texttt{PYTERPOL} fit of the spectrum from RJD~60793.672, observed near one elongation of the Aa-Ab binary.}
\label{spectra_fit}
\end{figure}

We used the program \texttt{PYTERPOL}, developed by J. Nemravová \citep{Nemravova2016, Nasseri2014}, which allows the determination of radiative parameters and individual RVs of a multiple star system using the comparison of real and synthetic spectra.
\texttt{PYTERPOL} was applied to all available blue segments of CTIO spectra, over the
wavelength range 4630 to 5428~\AA\,, to estimate the radiative properties and relative luminosities of the three
stellar components Aa, Ab, and Ba present in the spectra. The results are summarised in Table~\ref{pyter}, and an example of the fit for one spectrum  close to the elongation of the Aa-Ab binary is shown in Fig.~\ref{spectra_fit}. The program also returns the RVs of the components. However, we do not use them in the following analyses, since in this particular case, three relatively similar spectra
and a weak tertiary are not ideal, although they essentially confirm the preliminary results from direct measurements in \texttt{reSPEFO} for the Aa primary and Ab secondary. The disentangled spectra and RVs from \texttt{KOREL} give more accurate results for all 3 components Aa, Ab, and Ba, and the line-by-line analysis also provides some estimates of their errors for the following analyses in Sects. \ref{Eclipsing binary B} and \ref{Ellipsoidal binary A with an eccentric orbit}.

\begin{table}[h]
\centering  
\caption{Properties of individual stellar components from \texttt{PYTERPOL}.} 
\label{pyter}              
\begin{tabular}{l l l l}    
\hline\hline \noalign{\smallskip}           
Parameter & Aa & Ab & Ba\\   
\hline \noalign{\smallskip}
$T_\mathrm{eff}$ (K) & $31700\pm300$ & $22000\pm1000$ & $16070\pm500$ \\
$\log g$ & $3.87\pm0.10$ & $3.92\pm0.15$ & $3.27\pm0.10$ \\
$v_\mathrm{rot} \sin i$ & $108\pm5$ & $141\pm10$ & $250\pm10$ \\
(\ks)&&&\\
$L$  &$0.597\pm0.010$ & $0.316\pm0.010$ & $0.084\pm0.010$\\
\hline                    
\end{tabular}
\tablefoot{Obtained from fitting the blue CTIO spectra over the
wavelength interval 4630 -- 5428~\AA. $L$ denotes relative luminosity.}
\end{table}

\begin{figure}[t]
\resizebox{\hsize}{!}{\includegraphics[angle=0]{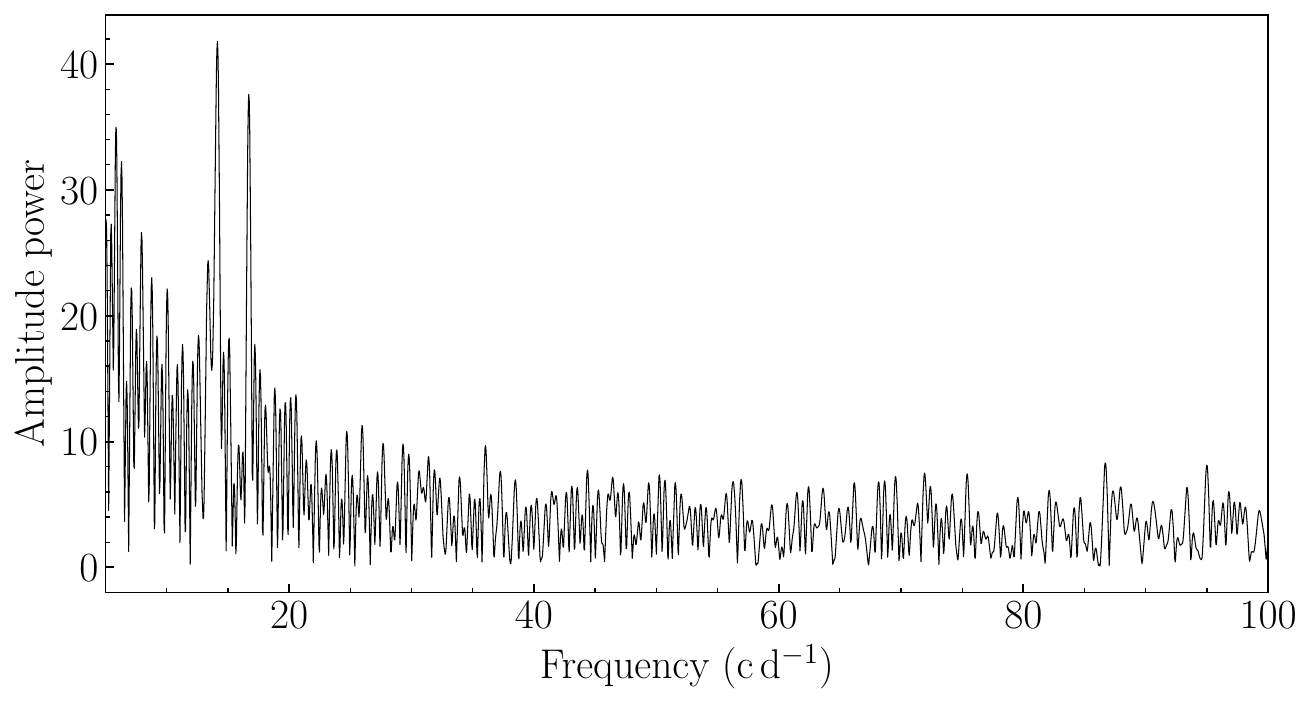}}
\caption{Amplitude periodogram of TESS data from RJD 60070.22-60072.61.  The highest peak at frequency of 14.14~\cd\ corresponds to period  0\fd071. The second highest peak at frequency of 16.70~\cd\ corresponds to period 0\fd060.}
\label{periodogram_oscilations}
\end{figure}

\section{Analyses of TESS light curves}
\begin{figure*}[h]
\centering
\resizebox{\hsize}{!}
{\includegraphics[angle=0]{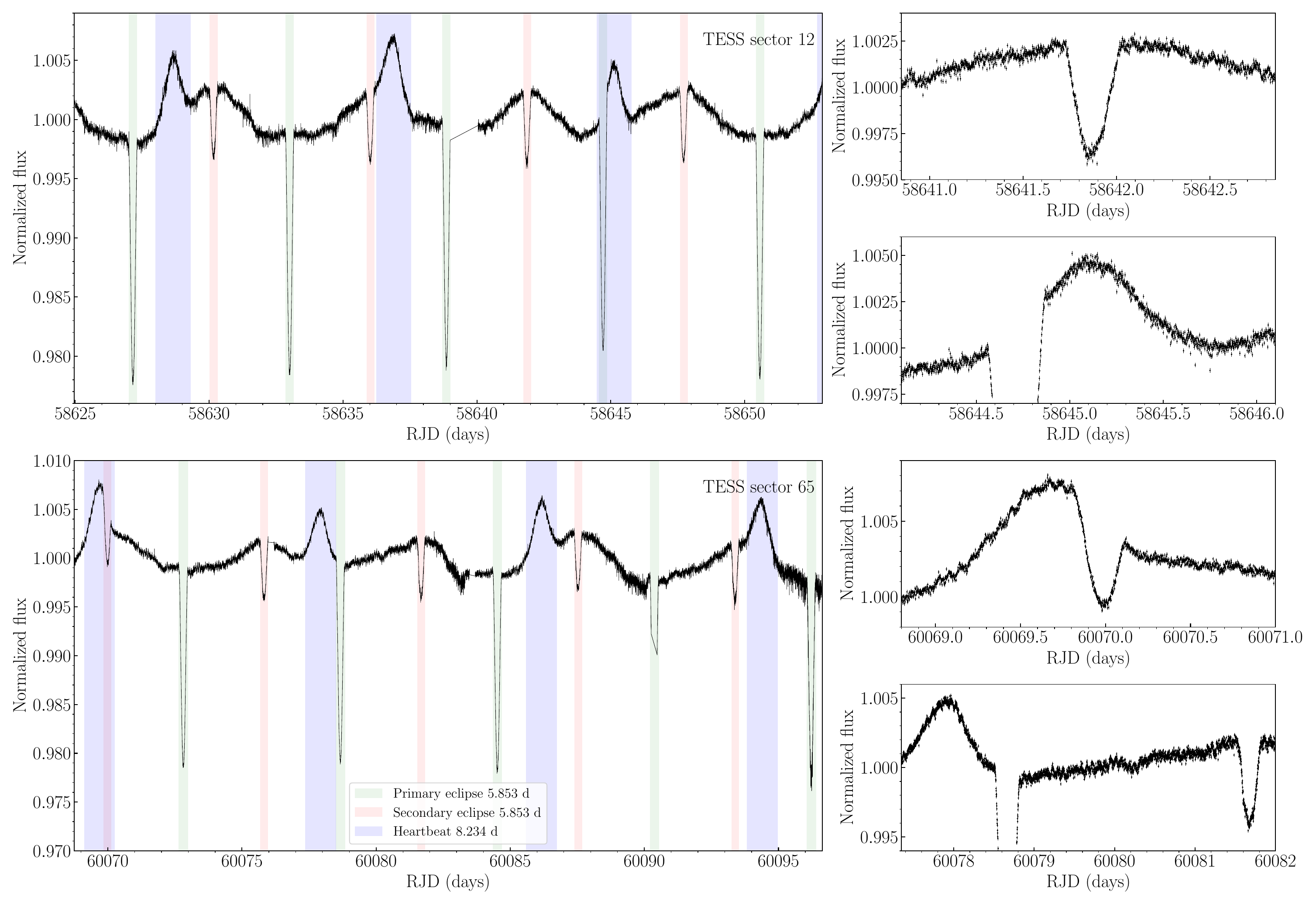}}
\caption{Left: TESS light curve of \hd from sector 12 (top) and 65 (bottom) with the 120 s exposure times
using SAP flux. Primary (green) and secondary eclipses (red) of binary B with a 5\fd853 period and heartbeat effect of binary A (blue) with a 8\fd234 period are shown. Primary eclipse at RJD 58644.71 and secondary eclipse at RJD 60069.69 of binary B is reduced by heartbeat of binary A. Right: Zoomed plots of visible oscillations with a period of 0\fd071. This small variability is not always clearly visible due to changing signal noise. LCs are normalized per sector to their median values.}
\label{LC_sector_65}
\end{figure*}

\begin{figure*}[t]
\centering
    \centering
    \includegraphics[width=0.49\linewidth]{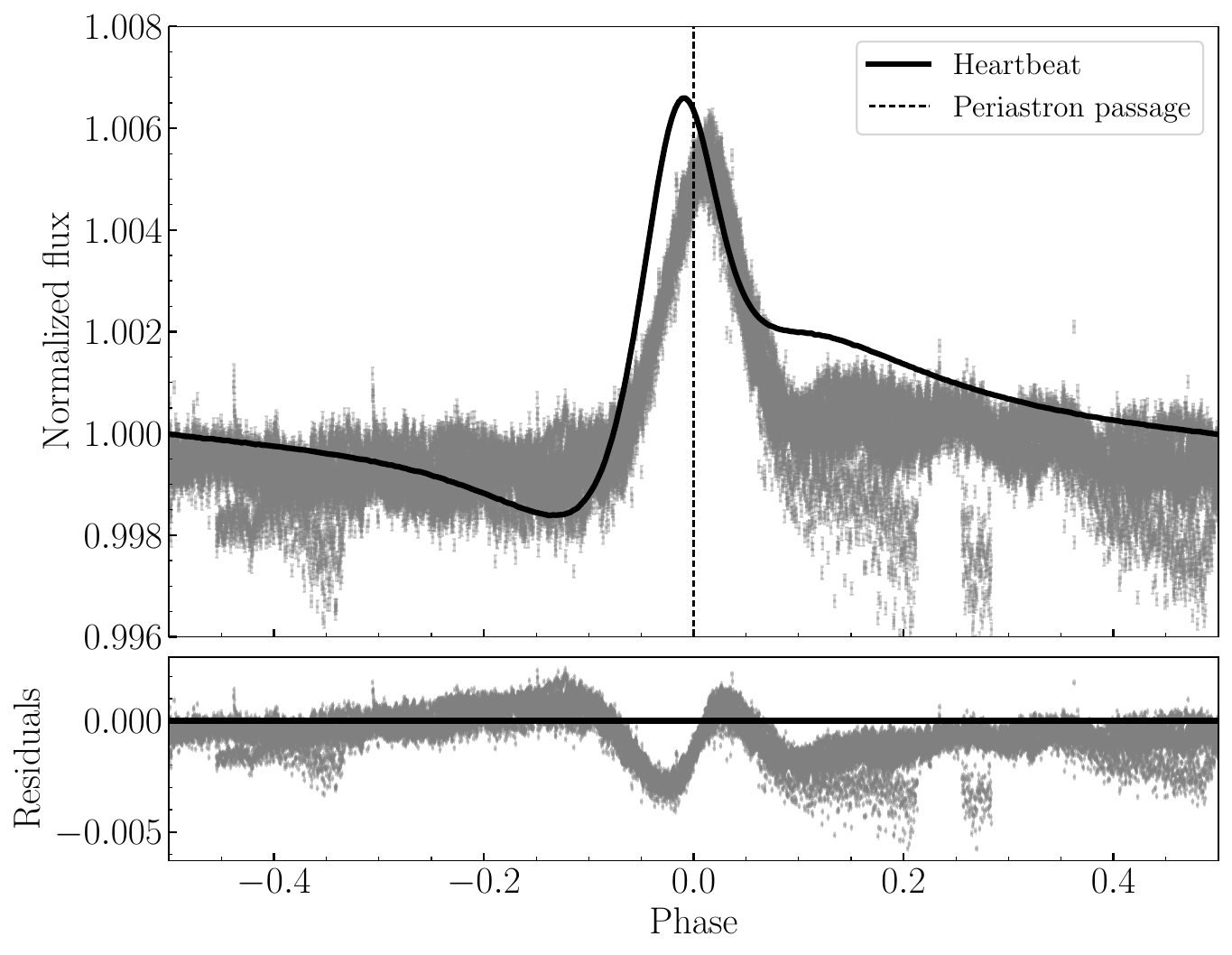}
    \includegraphics[width=0.49\linewidth]{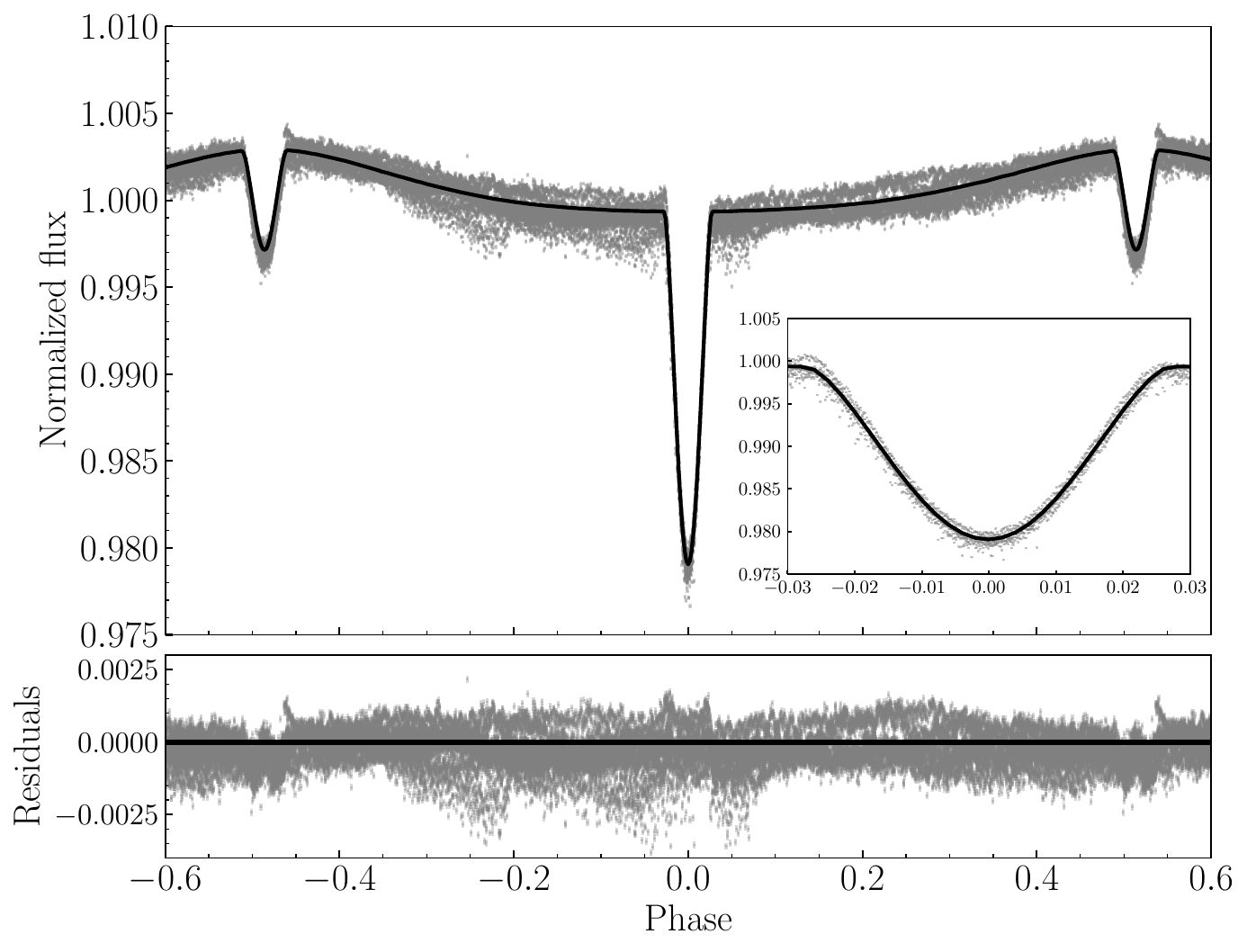}
\caption{Left: Phase folded flux of binary A after subtraction of eclipsing model for ephemeris $T_\textrm{0, perpass}=\textrm{RJD } 59995.6424 + 8.23459\times E $. Maximum of flux occurs shortly after the periastron passage. Right: Light curve of binary B from TESS without heartbeat from binary A. Ephemeris of $T_\textrm{0, supconj}=\textrm{RJD } 58627.15594 + 5.8528657\times E$ is used. Both available sectors 12 and 65 were used in all plots and are normalized to the median value per sector.}
    \label{LCs_phased}
\end{figure*}

An inspection of the TESS light curves of \hd from sectors 12 and 65 with 120 s exposure times, with both the ellipsoidal (heartbeat) pair A in eccentric orbit and the eclipsing pair B visible (see Fig.~\ref{LC_sector_65}), shows immediately that there are brightenings in the light curve
that repeat periodically with the 8\fd 234 period, and their positions correspond to instants near the periastron
passages of subsystem A. This effect is now known as the 'heartbeat effect'. Phase plots of TESS photometry from both sectors with the ephemeris of the final orbital solution for
both subsystems A and B, are shown in Fig.~\ref{LCs_phased}.

Small periodic flux variations are observed throughout the orbit. We have carried out a~period search on parts of the TESS data without the eclipses and brightenings near the periastron passages in the 8\fd234 orbit. The corresponding periodogram can be seen
in Fig. \ref{periodogram_oscilations} and shows two distinct peaks at 14.14~\cd, and 16.70~\cd, corresponding to periods of 0\fd071, and 0\fd060. However, inspection of the data plots versus time shows that the amplitude of these changes varies with time. No simple clear periodicity is 
obvious on the level of accuracy of the TESS data.

\subsection{Eclipsing binary B}
\label{Eclipsing binary B}

To obtain LC of eclipsing binary B, we manually removed data points related to 8\fd234 brightenings of star~A, shown in Fig. \ref{LC_sector_65} for both available normalised TESS sectors 12 and 65. The 'clean' normalised flux of 
binary B, phased with a period of 5\fd853, can be seen in Fig. \ref{LCs_phased} right. We have used \texttt{PHOEBE\,2}\footnote{\url{https://phoebe-project.org}} (Physics of Eclipsing Binaries; \citealt{Prsa2016}) for further modelling of the radial and light curves of the system~B. 

After preliminary trials with different parameters, we used the Markov Chain Monte Carlo method (MCMC) with 10000 steps and 95 walkers to identify possible degeneracies and probe the probability space. MCMC was re-run multiple times with slightly different initial values to be sure that we properly covered the global minimum of the solution. As for the preliminary \fotel solutions, we derived local systemic velocities for the \korel RVs of component Ba
and pre-whitened these RVs for the long orbit for use in \texttt{PHOEBE\,2}.
Using radiative properties for the component Ba from \texttt{PYTERPOL} and TESS LC, we have derived the elements listed in Table \ref{orbital_parameters}, assuming that the orbital period $P_{\scriptscriptstyle\mathrm{B}}$ remains constant. We note that some of the tabulated errors might be underestimated. We used a third light fraction of 0.85, as the value computed by \texttt{PYTERPOL} does not converge to a reasonable solution. For star Ba, we had to use the logarithmic limb darkening law and fit its coefficients, since the interpolated limb-darkening law is not yet available for hot stars in \texttt{PHOEBE\,2}. 
We note in passing that if the component Ba is a main-sequence star, the radius of Bb, which follows from the solution, is too large for a main-sequence star with \tef $\approx$ 6000~K and implies that the component Bb is an~evolved, giant star.

\subsection{Ellipsoidal binary A with an eccentric orbit}
\label{Ellipsoidal binary A with an eccentric orbit}

As obtaining separate LC and fitting radii of binary A is much more difficult, we decided to first probe the orbital parameter space using the MCMC method with 10000 steps and 95 walkers, and derived the final orbital solution for binary A based only on the radial velocities. The results of the final orbital solution are in Table~\ref{orbital_parameters}.

A possible way to obtain the LC of the ellipsoidal binary A is to subtract the synthetic LC of binary B from the original TESS data. As data are normalised to 1, we subtracted the model from the real data and added 1. This can be seen in Fig. \ref{LCs_phased} left. It would be theoretically possible to use both RV and TESS flux to model the system for binaries A and B separately. 
We have assumed blackbody atmospheres and temperatures of $T_{\scriptscriptstyle\mathrm{Aa}}=31700$ K and  $T_{\scriptscriptstyle\mathrm{Ab}}=22000$ K from \texttt{PYTERPOL}, and assumed that both Aa and Ab are main-sequence stars. Using estimates of radii $R_{\scriptscriptstyle\mathrm{Aa}}=6.8\ \mathcal{R}_\odot$, $R_{\scriptscriptstyle\mathrm{Ab}}=4.3\ \mathcal{R}_\odot$,  and masses $M_{\scriptscriptstyle\mathrm{Aa}}=17\ \mathcal{M}_\odot$, $M_{\scriptscriptstyle\mathrm{Ab}}=13.6\ \mathcal{M}_\odot$ based on \citet{mr88}, which leads to an inclination of $i=57^\circ$, 
we have derived a solution in Fig. \ref{LCs_phased} left. 

However, in the case of HD 135160, we point out that the shape of the brightened light curve is slightly different from the usual heartbeat. This is shown by the disparity between the data and the model in \texttt{PHOEBE\,2} (Fig. \ref{LCs_phased} left), which cannot explain its shape. As no  reliable estimators for modelling heartbeat light curves are available, we tried multiple combinations of different parameters, changing radii, inclination, and rotational periods of stars, but none of these models came close to the observed shape of the heartbeat. Even changes in temperature did not change the shape of the light curve.

We also tried to obtain orbital parameters, including the inclination of the LC of binary A using the model by \cite{Kumar1995ApJ}, with no success as the model required $i\approx30^\circ$, which would lead to unphysical masses of stars Aa and Ab. We can only speculate that, with high-mass stars, other secondary effects are present during the periastron. Contrary to the \texttt{PHOEBE\,2} model, real stars have finite response times to tidal forces, which might cause a delay in maximum flux. If these estimates of masses based on main sequence stars are reliable, binary A will lead to a total mass of the system on the high end of currently discovered heartbeat stars.

\begin{table}
\begin{flushleft}
\caption{Derived parameters of eccentric ellipsoidal A and eclipsing binary B, using \texttt{PHOEBE\,2} MCMC.}
\label{orbital_parameters}
\begin{tabular}{lcc}
\hline\hline\noalign{\smallskip}
    &A&B\\
    &heartbeat binary&eclipsing binary\\
\hline\noalign{\smallskip}
$P$ (d)           & $8.23459^{+0.00003}_{-0.00002}$&$5.8528657^{+0.0000008}_{-0.0000007}$  \\
$T_\textrm{0, perpass}$ (RJD)& $59995.6424^{+0.0047}_{-0.0046}$&-- \\
$T_\textrm{0, supconj}$ (RJD)& --& $58627.15594^{+0.00020}_{-0.00014}$\\
$e$ & $0.4236^{+0.0015}_{-0.0015}$& $0.1081^{+0.0008}_{-0.0008}$\\
$q$ & $0.8023^{+0.0031}_{-0.0032}$& --\\
$i$ (\textdegree) & -- & $81.92^{+0.02}_{-0.02}$\\
$\omega$ (\textdegree)   & $203.38^{+0.25}_{-0.25}$  & $281.5^{+0.1}_{-0.1}$ \\
$a\sin{i}$ ($\mathcal{R}_\odot$)& $47.953^{+0.082}_{-0.083}$  &--\\
$a_{\scriptscriptstyle\mathrm{a}}\sin i$ ($\mathcal{R}_\odot$)  & --& $3.446^{+0.013}_{-0.014}$ \\
$T_{\scriptscriptstyle\mathrm{a}}$ (K)& 31700$^*$&  16070$^*$ \\
$T_{\scriptscriptstyle\mathrm{b}}$ (K)& 22000$^*$&  $6000^{+500}_{-500}$  \\
$T_{\scriptscriptstyle\mathrm{b}}/T_{\scriptscriptstyle\mathrm{a}}$ &--& $0.3718^{+0.0002}_{-0.0002}$ \\
$R_{\scriptscriptstyle\mathrm{b}}/R_{\scriptscriptstyle\mathrm{a}}$ &--& $0.8771^{+0.002}_{-0.002}$ \\
$(R_{\scriptscriptstyle\mathrm{a}} + R_{\scriptscriptstyle\mathrm{b}})/a$              &--& $0.2172^{+0.0002}_{-0.0003}$ \\
ld coeff1$^{**}$                    &--& $0.294^{+0.012}_{-0.009}$ \\
ld coeff2$^{**}$                   &--& $-0.049^{+0.055}_{-0.061}$ \\
third light &  --& 0.85 fix \\
\hline\noalign{\smallskip}
\end{tabular}
\end{flushleft}
\tablefoot{$^*$ Fixed at the values from \texttt{PYTERPOL}.
 $^{**}$ Component Ba.} 
\end{table}

\section{Discussion and conclusions}

We have presented a detailed study of a newly discovered massive 2+2 quadruple system HD~135160, with a unique architecture consisting of an eccentric ellipsoidal (heartbeat) binary and an eclipsing binary pair, based on our four-year systematic echelle spectroscopy and photometry from TESS. We demonstrate orbital motion of pair~A on a common orbit, with a period between 1600 and 2200 days (4.4 - 6 years).
Three components, Aa, Ab, and Ba, are massive B-type stars with temperatures of $31700\pm300, 22000\pm1000$, and $16070\pm500$ K, respectively, while the fourth star is probably of a spectral type F with a temperature of $6000\pm500$ K. Short and shallow eclipses from subsystem B are slightly more than 2 \% of total flux, which is explained by the fact that binary B corresponds to less than 15 \% of total flux from the system. The huge difference in the temperatures of Ba and Bb results in a strong reflection effect. 

We also found small cyclic light variations on a time scale of 0\fd071 (14.14~\cd), with a~variable amplitude. This could indicate a beat of two close
periods, but we are unable to study these changes in more detail on the level of accuracy of the TESS data. We note that component Aa lies within the $\beta$~Cephei instability strip where stellar pulsations are expected. Although the observed period of 0\fd071 is shorter than typical $\beta$~Cep periods (0.1–0.3 d), such a short period is not entirely unprecedented \citep{Stankov_2005}. Additionally, tidal distortions in this extreme heartbeat system can, in principle, modify the stellar mode cavity and affect pulsation frequencies \citep{Fuller2020, Fuller2025}. The nature of the observed 0\fd071 variation therefore remains unclear and warrants further investigation.

There is no evidence of \ha emission in any of our spectra; we suspect the earlier classification of the object as a Be star resulted from misinterpreting double absorption lines, which are occasionally present, as a central emission line.

As shown by the precise radial velocities obtained over four years, subsystem A is on a highly eccentric orbit, which causes periodic brightenings slightly after the periastron passage. Contrary to numerous other heartbeat stars, we are unable to obtain the inclination (and masses) of the subsystem, as the shape of the brightened peak is not easily modelled. We suggest that the brightenings may not be caused by extreme tidal distortions near the periastron, but that other secondary effects may cause a distinct shape of the light curve.
We modelled both binary pairs separately, neglecting any mutual interactions such as light-travel effects or eclipse timing variations. 

We hope that future spectral coverage of the entire orbit and dedicated interferometric observations might lead to the determination of precise radial velocities of components Aa, Ab, and Ba, and the orbital parameters of the long orbit. It will then be possible to investigate the evolutionary stage that leads to this remarkable configuration.

\section*{Data availability}
Table 2 is available only in electronic form at the CDS via anonymous ftp to cdarc.u-strasbg.fr (130.79.128.5) or via http://cdsweb.u-strasbg.fr/cgi-bin/qcat?J/A+A/.

\begin{acknowledgements}
This study is based on the new spectroscopic observations with the CHIRON echelle spectrograph
of the Cerro Tololo Inter-American Observatory (CTIO) and on space photometry from
the Transiting Exoplanet Survey Satellite (TESS) which are publicly available from the Mikulski Archive for
Space Telescopes (MAST). Funding for the TESS mission is provided by NASA’s Science Mission Directorate.

We gratefully acknowledge the use of the latest publicly available version
of the programs \fotele\ and \korele, written by P.~Hadrava, the reduction program for spectroscopic reductions \respefoe,
written by A.~Harmanec, and the program for spectra modelling written by J.~Nemravov\'a. M. Zummer thanks M. Wrona for advice with this paper, A. Prša and K. Hambleton Prša for advice with \texttt{PHOEBE\,2} modelling. 

Finally, we acknowledge the use of the electronic database from
the CDS, Strasbourg, and the electronic bibliography maintained by
the NASA/ADS system.
\end{acknowledgements}

\bibliographystyle{aa}
\bibliography{hd}

\end{document}